\documentclass[prd,showpacs,preprintnumbers,twocolumn,amsmath,nofootinbib,amssymb]{revtex4}
\usepackage{graphicx,color,dcolumn,booktabs,bm}
\usepackage{longtable,lscape}
\usepackage{txfonts}
\usepackage{overpic}
\usepackage{amssymb}
\usepackage{epstopdf}
\usepackage{indentfirst}
\usepackage{feynmf}   %{feynmp}
\usepackage{slashed}  %for Feynman symbols
\usepackage{cases}
\usepackage{float}
\usepackage{rotating}
\usepackage{multirow}
\usepackage{array}
\usepackage{booktabs}
\usepackage{epsfig,dsfont,amssymb,amsmath,amsfonts,amsbsy,mathrsfs}
\graphicspath{{Figures/}} %
\usepackage[colorlinks, citecolor=blue,anchorcolor=red,menucolor=red, linkcolor=red,filecolor=red,runcolor=red,urlcolor=blue,frenchlinks=red]{hyperref}
\makeatletter
\@addtoreset{equation}{section}
\makeatother

\allowdisplaybreaks

\begin{document}
\title{Study of Unflavored Light Mesons with $J^{PC}=2^{--}$}

\author{Dan Guo$^{1,3}$}\email{guod13@lzu.edu.cn}
\author{Cheng-Qun Pang$^{2}$}\email{xuehua45@163.com}
\author{Zhan-Wei Liu$^{1,3}$}\email{liuzhanwei@lzu.edu.cn}
\author{Xiang Liu$^{1,3}$}\email{xiangliu@lzu.edu.cn}
\affiliation{
$^1$School of Physical Science and Technology, Lanzhou University, Lanzhou 730000, China\\
$^2$College of Physics and Electronic Information Engineering, Qinghai Normal University, Xining 810000, China\\
$^3$Research Center for Hadron and CSR Physics, Lanzhou University and Institute of Modern Physics of CAS, Lanzhou 730000, China
}
\begin{abstract}

The  unflavored light meson families, namely $\omega_2$, $\rho_2$, and $\phi_2$, are studied systematically by  investigating the spectrum and the two-body strong decays allowed by Okubo-Zweig-Iizuka rule.
Including the four experimentally observed states and other predicted states, phenomenological analysis of the partial decay widths can verify the corresponding assignments of these states into the families.
Moreover, we provide typical branching ratios of the dominant decay channels, especially for missing ground states, which is helpful to search for or confirm them and explore more properties of these families at experiment.

\end{abstract}


\maketitle

%====================================================================================================================================

\section{introduction}\label{sec1}
Since Quark model had been proposed in 1964 \cite{GellMann:1964nj,Zweig:1964jf}, big progresses on the study of light hadron spectrum have been made by joint effort from both theorists and experimentalists \cite{Tanabashi:2018oca}.
Among the light hadron families, light mesons are an indispensable part and still are unclear, which is the main reason why investigating and exploring light mesons becomes an interesting physical aim of the present running experiments like BESIII and COMPASS, and forthcoming experiments like GlueX and PANDA.

In the past years, there exist abundant phenomenological works involved in pesudoscalar mesons associated with these observed $X(1835)$, $X(2120)$, $X(2370)$, and $X(2500)$ \cite{Yu:2011ta,Wang:2017iai}, vector mesons with the reported $Y(2175)$ \cite{Wang:2012wa,Piotrowska:2017rgt}, $\rho$ and $\rho_3$ mesons \cite{Piotrowska:2017rgt,He:2013ttg}, axial vector mesons \cite{Chen:2015iqa,Giacosa:2017pos}, tensor mesons \cite{Ye:2012gu,Giacosa:2005bw,Pang:2014laa}, pseudotensor mesons \cite{Wang:2014sea,Koenigstein:2016tjw}, kaons \cite{Pang:2017dlw}, and higher spin mesons \cite{Pang:2015eha}.
These studies have played crucial role not only to establish light meson spectrum, but also enlarge our understanding to these new hadron states like $X(1835)$ and $Y(2175)$.

When further checking the experimental status of light mesons \cite{Tanabashi:2018oca}, we notice an interesting phenomenon, i.e., the ground states of unflavored light meson families with $J^{PC}=2^{--}$ are still absent, their higher excited states are not well established yet, and they are only collected into \emph{Further States} in Particle Date Group (PDG) \cite{Tanabashi:2018oca}.
This phenomenon stimulates our interests in exploring unflavored light mesons with $J^{PC}=2^{--}$ thoroughly.

According to the isospin, unflavored light mesons with $J^{PC}=2^{--}$ can be categorized into three groups, which are isovector $\rho_2$ meson family, and isoscalar $\omega_2$ and $\phi_2$ meson families.
Though experimental information of $2^{--}$ unflavored light ground meson is totally barren, their counterparts $1^{--}$ and $3^{--}$ meson are properly recognized.
By comparing the similarity of them, the mass spectrum of $\omega_2$, $\rho_2$ and $\phi_2$ ground states is successfully constructed.
In addition, we predict the behaviors of two-body strong decays allowed by Okuba-Zweig-Iizuka (OZI) rule, which are the key information of explaining the disappearance reason for them.
Here, potential decay channels and typical branching ratios which are valuable to explore in future experiment are also provided.
On the other hand, combining with the theoretical investigation of $K^*_2$ meson \cite{Pang:2015eha}, we can found an integrated nonet, and then systematically compare with $1^{--}$ and $3^{--}$ light mesons.

In PDG \cite{Tanabashi:2018oca}, we can find $\rho_2(1940)$, $\omega_2(1975)$, $\omega_2(2195)$, and $\rho_2(2225)$ as further states in $p\bar{p}$ reaction.
Based on the experimental information of these four $2^{--}$ states, we continue to discuss their possible assignments into $\rho_2$  and $\omega_2$ meson families with the mass spectrum analysis and two-body OZI-allowed strong decay calculation.
Besides, we predict the properties of the corresponding undiscovered $\phi_2$ mesons.
Additionally, masses and decay behaviors of third radial excitations are also calculated based on previous results.
We hope that our effort will be helpful to establish $\rho_2$, $\omega_2$ and $\phi_2$ meson families.

Comparing with other theoretical calculations \cite{Barnes:1996ff,Godfrey:1998pd,Ebert:2009ub}, not only a more detailed analysis of the missing ground states but also a theoretical study of newly observed four resonances are included in this work.
Moreover, based on the consistence of experimental information, we predict the decay properties of the third excitation states.

We organize this paper as follows.
In Sec. \ref{sec2}, the mass spectrum of unflavored light meson is studied, where one can fix the mass range of ground state in unflavored light meson families with $J^{PC}=2^{--}$ and give the possible assignments of radial excitation states.
In Sec. \ref{sec3}, we first introduce the Quark-Pair-Creation (QPC) model, and then we discuss the strong decay behaviors of these discussed states with fixed mass spectrum.
In the end, Sec. \ref{sec4} is devoted into the summary and discussion.

%====================================================================================================================================
\section{analysis of mass spectrum}\label{sec2}
%====================================================================================================================================

\begin{figure}[htbp]
  %\centering
  \flushleft
  \includegraphics[width=0.48\textwidth]{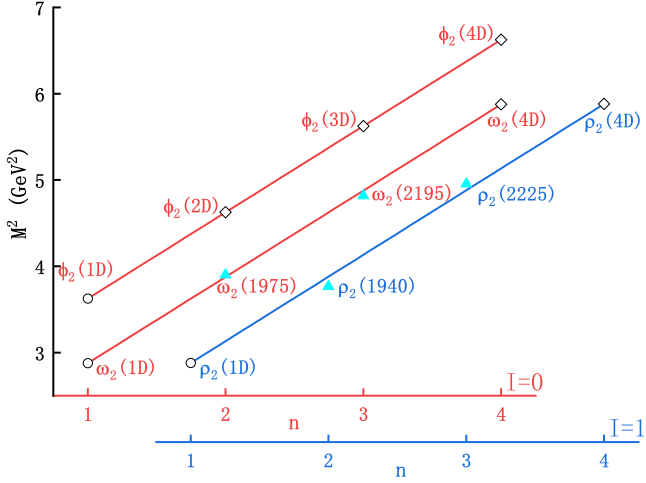}
  \caption{Regge trajectories of $\omega_2$, $\rho_2$ and $\phi_2$ families. Slope $\mu^2$ of trajectories is 1.0 $\rm{GeV}^2$. Here, the empty circles (or diamonds) represent the theoretical values of ground states (or radical excitations) calculated by GI model (or Regge trajectory). The cyan triangles denote the experiment data.}
  \label{reggefig}
\end{figure}

Although the light isoscalar and isovector states with $^{2S+1}L_J$ = $^{3}D_1$ or $^3D_3$ are relatively well established and categorized because of many discovered candidates  \cite{Wang:2012wa,He:2013ttg}, there are only four unflavored light $^3D_2$ mesons discovered at experiment.
By using the Crystal Barrel detector's data, four $2^{--}$ resonances are reported for $p\bar{p}$ collision in 2002 \cite{Anisovich:2002su,Anisovich:2011sva}, whose masses and widths are listed
\begin{eqnarray}
  \rho_2(1940):\quad m=&1940\pm40 \;\mathrm{MeV},\quad \Gamma=&155\pm40 \;\mathrm{MeV}; \nonumber\\
  \rho_2(2225):\quad m=&2225\pm35 \;\mathrm{MeV},\quad \Gamma=&335^{+100}_{-50} \;\mathrm{MeV}; \nonumber\\
  \omega_2(1975):\quad m=&1975\pm20 \;\mathrm{MeV},\quad \Gamma=&175\pm25 \;\mathrm{MeV}; \nonumber\\
  \omega_2(2195):\quad m=&2195\pm30 \;\mathrm{MeV},\quad \Gamma=&225\pm40 \;\mathrm{MeV}. \nonumber
\end{eqnarray}

According to our previous works \cite{Wang:2012wa,He:2013ttg} and the review \emph{Quark Model} in PDG \cite{Tanabashi:2018oca}, the accompanying isoscalar $\omega(1650)$ and isovector $\rho(1700)$ are generally considered as $^3D_1$ ground states, and isoscalar $\omega_3(1670)$ and isovector $\rho_3(1690)$ as $^3D_3$ ground states.
Those are consistent with the rough estimations as follows.
Because of same spin angular momentum and orbital angular momentum, the masses of $\omega_2$ and $\rho_2$ ground states should be around 1.7 GeV by comparing masses of the $^3D_1$ state and $^3D_3$ state.
Similarly, the mass of $\phi_2$ ground state is about 1.9 GeV by comparing that of $\phi_3(1850)$.

In most situation, the Godfrey-Isgur (GI) model \cite{Godfrey:1985xj} is used for mass spectrum analysis and has achieved great success since proposed. Here, we employ this model to obtain masses of ground states.% with improved precision.

The Regge trajectory is another effective approach to quantitatively study the mass spectrum of the radial excited light mesons \cite{Chew:1962eu,Anisovich:2000kxa}. As to higher excitations, masses of GI model calculation generally are larger than experimental results because of the coupled channel effect and the relativity effect. Thus, we use the Regge trajectory to obtain the masses of the radical excitation states. A general expression for the Regge trajectory is
\begin{equation}
    M^2=M_0^2+(n-1)\mu^2,
\end{equation}
where $M_0$ is the mass of ground state, $M$ is the mass of state with radical excitation number $n$, and $\mu^2$ denotes the trajectory slope.

%By combining the analysis of ground states from GI model and the analysis of excitation states from Regge trajectory, 

A general mass assignment of unflavored light mesons with $^3D_2$ is shown in Fig. \ref{reggefig}, and the four experimental states are well-arranged as $n=2$ or $3$ state. Here, we need to specify that the masses of $\rho_2(1D)$, $\omega_2(1D)$ and $\phi_2(1D)$ are taken from the GI model calculation \cite{Godfrey:1985xj}. By combining these theoretical inputs and experimental data, we can construct three  Regge trajectories just shown in Fig. \ref{reggefig}, by which can further predict the masses of other missing states.
As shown in Fig. \ref{reggefig}, linear typical Regge trajectories of $\omega_2$ and $\rho_2$ family are observed, and we can obtain the slope $\mu^2 = 1.0~\rm{GeV}^2$, which is gotten by fitting the experimental and the GI ground states data.
Moreover, when we take the mass of the ground state $\phi_2(1D)$ from GI model, the masses of $\phi_2$ family can also be obtained with the extrapolation from the Regge trajectory using the same slope $\mu^2 = 1.0~\rm{GeV}^2$.

%Combining with the analysis of $K_2^*$ from our previous work \cite{Pang:2017dlw}, where the study of quantum number and mass assignments, spin-mixing and strong decay properties is performed, , 

Besides these $\omega_2$, $\rho_2$, $\phi_2$ mesons, there exist the corresponding $K_2$ partners. In our previous work \cite{Pang:2017dlw}, the resonance parameter and partial decay width of these $K_2^*$ mesons were calculated. An integrated nonet is established as shown in Fig. \ref{nonet} and thus we can have a complete comprehension of $2^{--}$ light mesons.
The masses of unflavored mesons with different radical number $n$ are summarized in Table \ref{reggemass}, and these masses are employed to the following study for the decay widths.
On the other hand, analysis of decay behaviors will examine the reasonability of our assignments.

\makeatletter\def\@captype{figure}\makeatother
 \begin{minipage}[b]{0.35\linewidth}
   \centering
   \includegraphics[width=\textwidth]{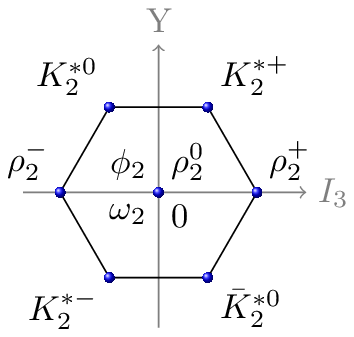}
   \caption{The $2^{--}$ nonet.}
   \label{nonet}
 \end{minipage}
 \quad
 \makeatletter\def\@captype{table}\makeatother
 \begin{minipage}[b]{0.5\linewidth}
 \centering
   \begin{tabular}{p{0.5cm}ccc}
     \hline
     \hline
     n & $\omega_2$ & $\rho_2$ & $\phi_2$ \\
     \hline
     1 & 1696 & 1696 & 1904 \\
     2 & 1975 & 1940 & 2151 \\
     3 & 2195 & 2225 & 2372 \\
     4 & 2424 & 2424 & 2574 \\
     \hline
   \end{tabular}
   \caption{The masses of $\omega_2$, $\rho_2$, and $\phi_2$ families in units of MeV.}
   \label{reggemass}
 \end{minipage}

Presently, in addition to the study of ground state masses from Godfrey and Isgur \cite{Godfrey:1985xj}, and Ebert {\it et al.} \cite{Ebert:2009ub}, there are also another two works \cite{Barnes:1996ff,Godfrey:1998pd} indicating ground state masses of $\omega_2$ and $\rho_2$ as $\sim$$1.7$ GeV comparable with ours.
Furthermore, Ebert {\it et al.} also predict the masses of the first and second excitations, and the masses of $\omega_2$ and $\rho_2$ families are generally consistent with ours, but the masses of $\phi_2$ excitations is higher than ours.
In the next section, we give a complete analysis of strong decay behaviors based on the well-established mass assignments.

%====================================================================================================================================
\section{strong decay behaviors}\label{sec3}
%====================================================================================================================================

In this section, we firstly introduce the phenomenological model applied to obtain the information for the OZI-allowed hadronic strong decays. We then illustrate strong decay behaviors of $\omega_2$, $\rho_2$ and $\phi_2$ families in details, respectively.

%====================================================================================================================================
\subsection{A Brief Introduction to the QPC Model}\label{QPCmodel}
%====================================================================================================================================

\begin{figure*}[htbp]
  \centering
  \includegraphics[width=0.8\textwidth]{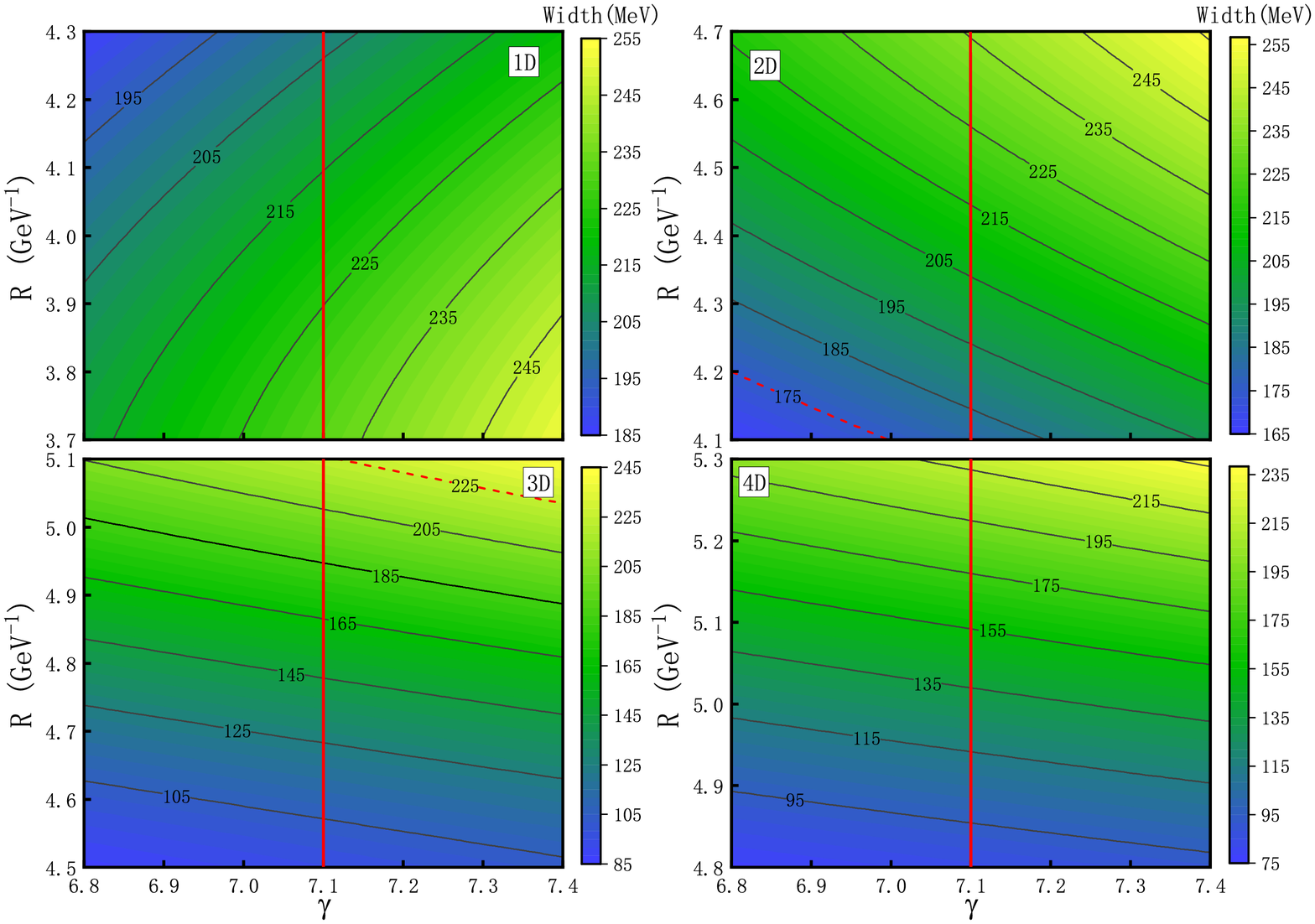}
  \caption{The total widths of $\omega_2$ family ($n=1$ to 4) with dependence of $R$ and $\gamma$.  Different colors correspond to different width values and the contour curves are marked with the width values. Specially, the red dashed contour lines means the experimental widths, and the red longitudinal solid lines depicts the $R$ dependence when $\gamma=7.1$.}
  \label{omegacontour}
\end{figure*}

The QPC model is proposed by Micu \cite{Micu:1968mk} firstly in 1968 and developed by Orsay group  \cite{LeYaouanc:1972vsx,LeYaouanc:1973ldf,LeYaouanc:1974cvx,LeYaouanc:1977fsz,LeYaouanc:1977gm}.
The QPC model assumes a pair of quark-antiquark $q\bar{q}$ is created from vacuum with $J^{PC}=0^{++}$ and then rearranged with the initial hadron to form two daughter hadrons.
For instance, the transition operator $\mathcal{T}$ of a meson decay progress $A\to B+C$ can be expressed as
\begin{eqnarray}
\mathcal{T}& = &-3\gamma \sum_{m}\langle 1m;1-m|00\rangle\int d^3 \mathbf{p}_3d^3 \mathbf{p}_4 \delta ^3 (\mathbf{p}_3+\mathbf{p}_4) \nonumber \\
&& \times \mathcal{Y}_{1m}\left( \frac{\mathbf{p}_3-\mathbf{p}_4}{2} \right) \chi _{1,-m}^{34} \phi _{0}^{34}
\omega_{0}^{34} b_{3}^{\dag} (\mathbf{p}_3) d_{4}^{\dag}(\mathbf{p}_4),\label{QPC1}
\end{eqnarray}
where $\mathcal{Y}_{lm}(\mathbf{p})=|\mathbf{p}|^lY_{lm}(\theta_p,\phi_p)$ is the solid spherical harmonic polynomial, and $\mathbf{p}_3$ and $\mathbf{p}_4$ depict momenta of quark and antiquark created from vacuum. $b_3^{\dag}$ ($d_4^{\dag}$) denotes quark (antiquark) creation operator.
$\chi_{1,-m}^{34}$, $\phi _{0}^{34}$, and $\omega_{0}^{34}$ denote spin triplet, flavor singlet, and color singlet wavefunctions, respectively.
The dimensionless parameter $\gamma$ describes the quark pair creation strength, which can be fitted by the experimental width data. From our fitted results of $2^{--}$ light states  with minimal $\chi^2=17.8$ $(\chi^2=\sum_{i}(\mathrm{Theo.}-\mathrm{Exp.})^2/\mathrm{Error}^2)$, $\gamma=7.1$ for the $u\bar{u}$ ($d\bar{d}$) pair creation, and the $s\bar{s}$ quark pair creation strength sets as $7.1/\sqrt{3}$ \cite{LeYaouanc:1977gm}.

Then the transition matrix of decay process in the the rest frame of particle $A$ reads as
\begin{eqnarray}
\langle BC | \mathcal{T}| A\rangle = \delta^3(\mathbf{P}_B+\mathbf{P}_C) \mathcal{M}^{M_{J_A}M_{J_B}M_{J_C}},\label{QPC2}
\end{eqnarray}
where $\mathbf{P}_B$ and $\mathbf{P}_b$ depict momenta of meson $B$ and $C$, respectively, and $\mathcal{M}^{M_{J_A}M_{J_B}M_{J_C}}$ is the decay amplitude with $M_{J_i}(i=A, B, C)$ describing the magnetic quantum number of the meson.

The helicity amplitude $\mathcal{M}^{M_{J_A}M_{J_B}M_{J_C}}$ can be converted into partial wave amplitude $\mathcal{M}^{JL}$ by the Jacob-Wick formula \cite{Jacob:1959at}, i.e.,
\begin{eqnarray}
\mathcal{M}^{JL}(\mathbf{P})&=&\frac{\sqrt{2L+1}}{2J_A+1}\sum_{M_{J_B}M_{J_C}}\langle L0;JM_{J_A}|J_AM_{J_A}\rangle \nonumber \\
&&\times \langle J_BM_{J_B};J_CM_{J_C}|{J_A}M_{J_A}\rangle \mathcal{M}^{M_{J_{A}}M_{J_B}M_{J_C}}.
\end{eqnarray}

To sum up, the general decay width writes as
\begin{eqnarray}
\Gamma_{BC}&=&\frac{\pi}{4} \frac{|\mathbf{P}|\mathcal{S}}{m_A^2}\sum_{J,L}|\mathcal{M}_{BC}^{JL}(\mathbf{P})|^2,
\end{eqnarray}
where $\mathcal{S}\equiv1/(1+\delta_{BC})$ denotes the statistic factor which is responsible for a situation if $B$ and $C$ are identical particles, and $m_A$ is the mass of initial particle.

Furthermore, the meson wavefunction is defined as mock state, i.e.
\begin{eqnarray}
&&|D(n^{2S+1}L_{JM_J})(\mathbf{P}_D)\rangle\nonumber\\
&&\quad\quad=\sqrt{2E_D}\sum\limits_{ {M_{S}},{M_{L}}}\langle LM_L;SM_{S}|JM_{J}\rangle \nonumber\quad\\
&&\quad\quad\times \int d^3\mathbf{p_D} \chi^{D}_{S,M_S}\phi^D \omega^D \Psi^D_{nLM_L}(\mathbf{p_D})\nonumber\\
&&\quad\quad\times|q_1(\tfrac{m_1}{m_1+m_2}\mathbf{P_D}+\mathbf{p_D}) \bar{q}_2(\tfrac{m_1}{m_1+m_2}\mathbf{P_D}-\mathbf{p_D})\rangle,
\end{eqnarray}
and here the spacial wavefunction $\Psi_{nLM_L}(\mathbf{p})$ of meson adopts the simple harmonic oscillator (SHO) wavefunction which has explicit form
\begin{eqnarray}
 &&\Psi_{nLM_L}(\mathbf{p})=R_{nL}(p, \beta)Y_{LM_L}(\Omega_p), \\
 &&R_{nL}(p,\beta)=\frac{(-1)^n(-i)^L}{ \beta ^{3/2}}e^{-\frac{p^2}{2 \beta ^2}}\sqrt{\frac{2n!}{\Gamma(n+L+3/2)}}{(\frac{p}{\beta})}^{L} \nonumber \\
 &&\quad\quad\quad\quad\times L_{n}^{L+1/2}(\frac{p^2}{ \beta ^2}),
\end{eqnarray}
where $Y_{LM_L}(\mathrm{\Omega})$ is spherical harmonic function, and $L_{n-1}^{L+1/2}(x)$ is the associated Laguerre polynomial.
The parameter $\beta=1/R$, and $R$ is obtained by reproducing the realistic root mean square radius via solving the Schr\"{o}dinger equation (see more details in Ref. \cite{Close:2005se}, though we proceed some improvements of the method and results).

In the following subsections, we perform phenomenological analysis of total widths and partial widths calculated by QPC model by comparing our results with experimental data. This analysis is helpful to explain why there does not exist any information of $2^{--}$ unflavored light ground states and reveal the underlying properties of these ground states and their radical excitations for more future experimental measurements, especially in BESIII detector, CMD-3, SND and KEDR detector.

%====================================================================================================================================
\subsection{$\omega_2$ Family}
%====================================================================================================================================

As shown in Fig. \ref{omegacontour}, the calculated total widths of $\omega_2$ family states are illustrated by contour lines which depend on $R$ and $\gamma$.
At present, experiment only observed $\omega_2(1975)$, $\omega_2(2195)$, $\rho_2(1940)$, $\rho_2(2225)$, where their widths were measured. Thus, we may take these experimental data to fix the $\gamma$ value to be $7.1$. Considering the uncertainty, we set $\gamma= 7.1\pm0.3$ in our concrete calculation. Generally, the total widths of $\omega_2(2D)$ and $\omega_2(3D)$ in our results agree well with those of $\omega_2(1975)$ and $\omega_2(2195)$ at experiment. It demonstrates the reasonability of setting the two states as $\omega_2(2D)$ and $\omega_2(3D)$ respectively, and also predicts there exists an undiscovered ground state.

\begin{figure}[tbhp]
  \centering
  \includegraphics[width=0.5\textwidth]{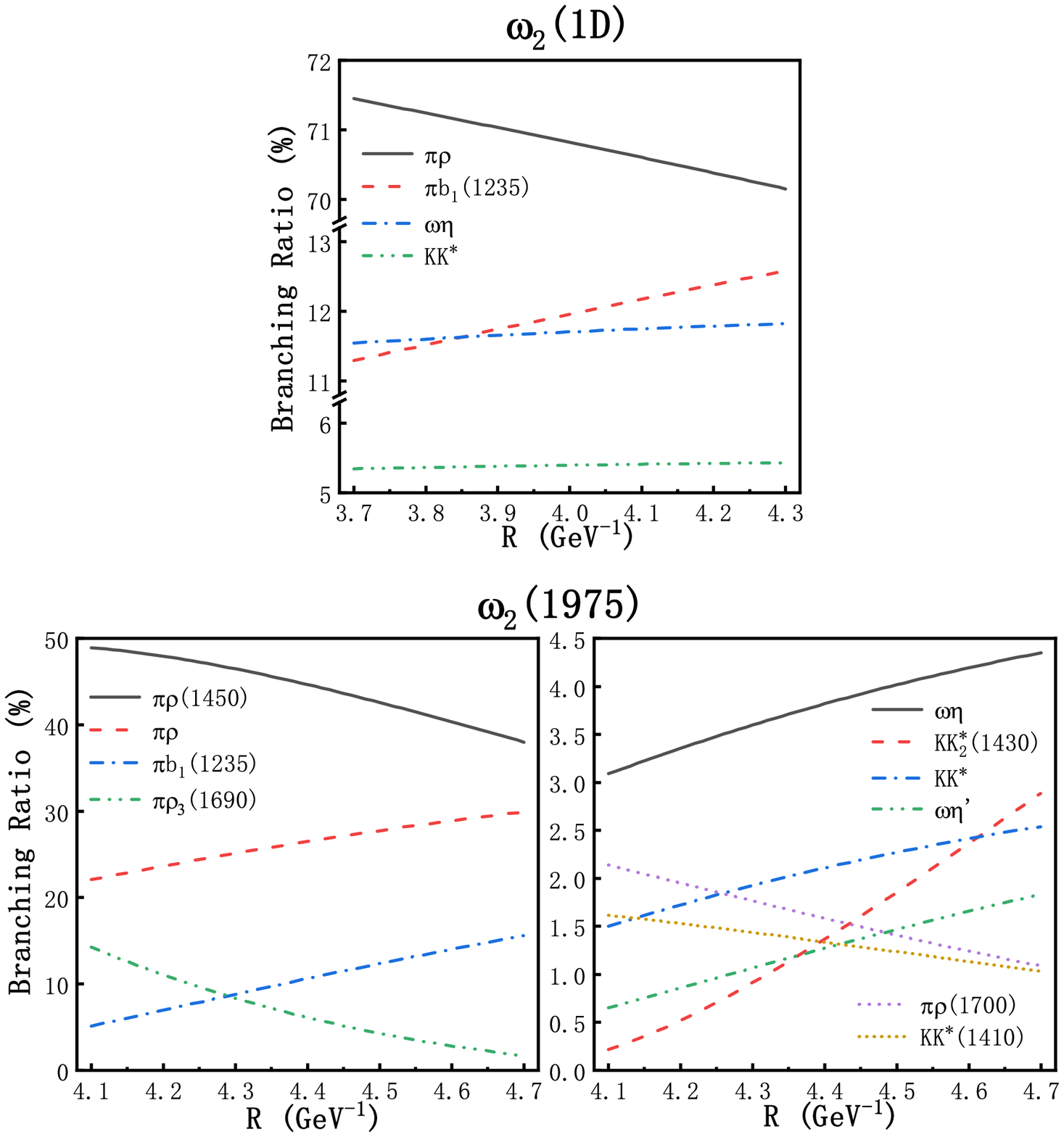}
  \caption{The branching ratios of $\omega_2(1D)$ and $\omega_2(1975)$ with $R$ dependence. Decay channels with branching ratios less than 1\% are neglected.}
  \label{omega12D}
\end{figure}

By assuming the ground state $\omega_2(1D)$ with mass equalling to 1696 MeV, the total decay width of missing $\omega_2(1D)$ reach up to 220 MeV when choosing fitted $\gamma=7.1$ and recommended $R$ value \cite{Close:2005se} shown in the top left of Fig. \ref{omegacontour}.
We present the branching ratios of $\omega_2(1D)$ at the top of Fig. \ref{omega12D}.
Corresponding partial decay width can be obtained by multiplying the total width and branching ratio.
The largest branching ratio comes from the $\pi\rho$ channel and is about $71\%$.
The $\pi b_1(1235)$, $\omega\eta$, and $KK^*$ channels are also important.
Besides, all branching ratios are not strongly dependent on $R$, and those of the $\pi b_1(1235)$ and $\omega\eta$ channels are close to each other.

The information of stable partial width ratios will be helpful for experimental searches, and we list them for $\omega_2(1D)$, $\Gamma(\pi\rho)/\Gamma(\pi b_1(1235)) = 5.58-6.33$, $\Gamma(\pi\rho)/\Gamma(\omega\eta) = 5.93-6.19$, $\Gamma(\pi\rho)/\Gamma(KK^*) = 12.9-13.4$, and $\Gamma(\omega\eta)/\Gamma(KK^*)=2.16-2.18$.
In Ref. \cite{Barnes:1996ff,Godfrey:1998pd}, the similar branching ratios of $\pi\rho$ are obtained, where the branching ratio of this $\pi\rho$ decay channel for $\omega_2(1D)$ is 74\% and 60\% given by Ref. \cite{Barnes:1996ff,Godfrey:1998pd}  respectively, and other branching ratios are also analogous.
As mentioned in Ref. \cite{Godfrey:1998pd}, $\pi_2(1670)$ is with the mass and total width similar to $\omega_2(1D)$, and also decays to $\pi\rho$, so that $\omega_2(1D)$ is possibly masked by $\pi_2(1670)$ in the $\pi\rho$ channel.
Therefore, $\omega\eta$ and $KK^*$ are the ideal channels suggested to search for $\omega_2(1D)$. The process $\omega_2(1D)\to b_1(1235)\pi\to \omega\pi\pi$ are also valuable since $\omega_2(1975)$ and $\omega_2(2195)$ are discovered in the $\omega\pi\pi$ channel.

The structure $\omega_2(1975)$ and $\omega_2(2195)$ are well determined by both $p\bar{p}\to\omega\eta$ and $p\bar{p}\to\omega\pi\pi$ processes with $\omega$ decaying to $\pi^+\pi^-\pi^0$ \cite{Anisovich:2011sva}.
The two resonances are classified as furthur states in PDG, which means these states are not confirmed well in experimental and more measurements are needed.
Thus, a detailed analysis of their categorization in nonet and decay properties are valuable to confirm them.

According to the analysis of Regge trajectory in previous section, $\omega_2(1975)$ is assigned as $n$ = 2, whose total width and branching ratios of calculated two-body decays are presented in the top right of Fig. \ref{omegacontour} and at the bottom of Fig. \ref{omega12D}, respectively.
The experimental total width of $\omega_2(1975)$ is 175$\pm$25 MeV, which is well reproduced in our results.

As to the partial widths of $\omega_2(1975)$, our calculation shows $\pi\rho(1450)$ are dominant decay channel and the branching ration reaches up to (38-49)\%.
$\pi\rho$, $\pi b_1(1235)$ and $\pi\rho_3(1690)$ are also main modes, and $\omega\eta$ is also considerable.
Corresponding to our calculation, $\pi\rho(1450)$, $\pi b_1(1235)$ and $\pi\rho_3(1690)$ modes can contribute to $\omega\pi\pi$ final state, and $\omega\eta$ mode also shows significant proportion.
It can explain why $\omega_2(1975)$ is discovered in $p\bar{p}\to\omega\eta$ and $p\bar{p}\to\omega\pi\pi$ processes by Crystal Barrel Collaboration.
Moreover, we can give partial decay width ratios which are stable against $R$ value, $\Gamma(\pi\rho(1450))/\Gamma(\pi\rho) = 1.27-2.22$, $\Gamma(\pi b_1(1235))/\Gamma(\pi\rho) = 0.23-0.52$, $\Gamma(\pi\rho_3(1690))/\Gamma(\pi b_1(1235)) = 0.11-2.76$, $\Gamma(\omega\eta)/\Gamma(\pi\rho) = 0.140-0.145$, $\Gamma(\omega\eta)/\Gamma(KK^*) = 1.71-2.06$, $\Gamma(\omega\eta)/\Gamma(\omega\eta') = 2.37-4.74$, and $\Gamma(\pi\rho(1700))/\Gamma(KK^*(1410)) = 1.05-1.32$.
Especially, the ratio $\Gamma(\omega\eta)/\Gamma(\omega\eta')$ is a promising issue to explore.
We believe more experimental tests of these ratios can examine the rationality of setting $\omega_2(1975)$ as $2^3D_2$ state.

\begin{figure}[!tbp]
  \centering
  \includegraphics[width=0.5\textwidth]{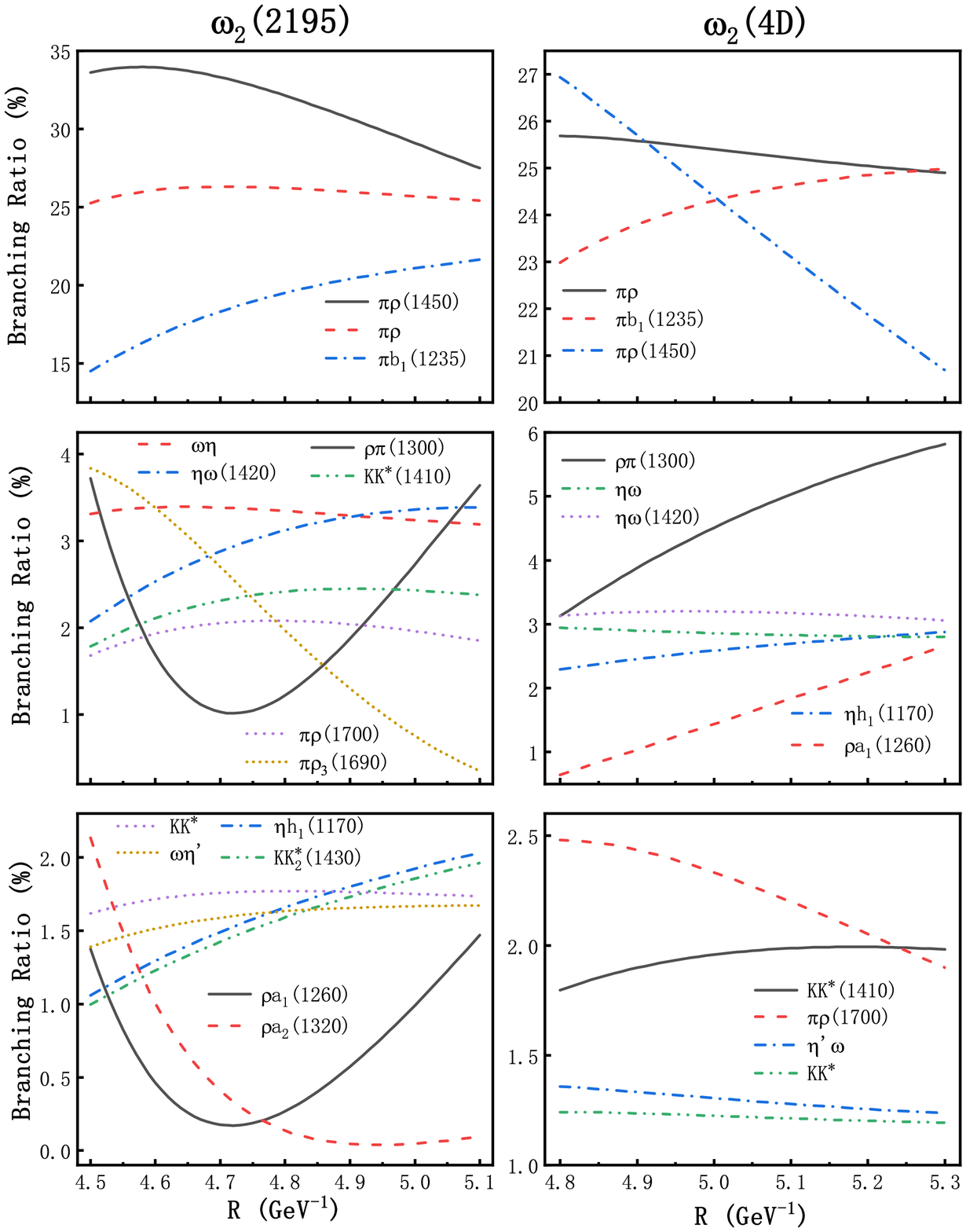}
  \caption{The branching ratios of $\omega_2(2195)$ and $\omega_2(4D)$ with $R$ dependence. Decay channels with branching ratios less than 1\% are neglected. }
  \label{omega34D}
\end{figure}

Another $J^{PC}=2^{--}$ state $\omega_2(2195)$ observed in Ref. \cite{Anisovich:2011sva} is categorized as $n$ = 3, and its experimental total width is 225$\pm$40 MeV.
A good agreement of our total width calculation with experimental data is observed in the lower left of Fig. \ref{omegacontour}.
The $R$ dependence of branching ratios for $\omega_2(3D)$ is illustrated in the left of Fig. \ref{omega34D}. The main decay channels are $\pi\rho(1450)$, $\pi\rho$, and $\pi b_1(1235)$ whose branching ratios are 27.5-33.6\%, 25.2-25.4\%, and 14.5-21.6\%, respectively.
$\pi\rho(1450)$ and $\pi b_1(1235)$ channels give contributions to $\omega\pi\pi$, and $\omega\eta$ also has a moderate percentage share, and thus it can be understood that $\omega_2(2195)$ is observed in these two final states. Here the decay width ratios weakly related to $R$ are listed, $\Gamma(\pi\rho(1450))/\Gamma(\pi\rho) = 1.08-1.33$, $\Gamma(\pi\rho)/\Gamma(\pi b_1(1235)) = 1.18-1.74$, $\Gamma(\pi\rho(1450))/\Gamma(\pi b_1(1235)) = 1.27-2.31$, $\Gamma(\omega\eta)/\Gamma(\omega(1420)\eta) = 0.94-1.59$, $\Gamma(\omega\eta)/\Gamma(KK^*(1410)) = 1.34-1.85$, $\Gamma(KK^*(1410))\Gamma(\pi\rho(1700)) = 1.06-1.28$, $\Gamma(\omega\eta)/\Gamma(KK^*) = 1.84-2.04$, $\Gamma(\omega\eta)/\Gamma(\omega\eta') = 1.91-2.38$, $\Gamma(KK^*)/\Gamma(\omega\eta') = 1.04-1.16$ and $\Gamma(\eta h_1(1170))/\Gamma(KK^*_2(1430)) = 1.03-1.06$.
Among these ratios, $\Gamma(\omega\eta)/\Gamma(\omega\eta')$ is suggested to study firstly at experiment.
Owing to the main channels contributing to $5\pi$ final state, these modes are inefficient to reconstruct at experiment.
Generally speaking, the total and partial widths depend more sharply on $R$ for higher radial excited states, and larger radial quantum number takes larger $R$ value.

Since no experimental signal of $\omega_2(4D)$, we predict its mass as 2424 MeV, and plot total width and branching ratios in the bottom-right corner of Fig. \ref{omegacontour} and the right column of Fig. \ref{omega34D}, respectively.
When choosing recommended $R$ and $\gamma$ value, the total width of $\omega_2(4D)$ is about 130 MeV.
Among three main decay channels, $\pi\rho(1450)$ linearly relies on $R$ value whose branching ratio reaches up to 20.7-26.9\%, while those of $\pi\rho$ and $\pi b_1(1235)$ are 24.9-25.7\% and 23.0-24.0\%, respectively.
Here, corresponding relative ratios are also given, $\Gamma(\pi\rho)/\Gamma(\pi b_1(1235)) = 1.00-1.12$, $\Gamma(\pi\rho)/\Gamma(\pi\rho(1450)) = 0.95-1.20$, $\Gamma(\pi b_1(1235))/\Gamma(\pi\rho(1450)) = 0.85-1.21$, $\Gamma(\rho\pi(1300))/\Gamma(\pi\rho) = 0.12-0.23$, $\Gamma(\omega\eta)/\Gamma(\eta h_1(1170)) = 0.97-1.29$, $\Gamma(\omega\eta)/\Gamma(\omega(1420)\eta) = 0.91-0.94$, $\Gamma(\omega\eta)/\Gamma(KK^*(1410)) = 1.41-1.64$, $\Gamma(\omega\eta)/\Gamma(\omega\eta') = 2.17-2.26$, and $\Gamma(\omega\eta')/\Gamma(KK^*) = 1.03-1.09$.
Advisable invariant mass spectra should be investigated at experiment are $\omega\pi\pi$, $\omega\eta$ and $KK^*$.
These decay information will be helpful for future experimental search.

%====================================================================================================================================
\subsection{$\rho_2$ Family}
%====================================================================================================================================

\begin{figure*}[htbp]
  \centering
  \includegraphics[width=0.8\textwidth]{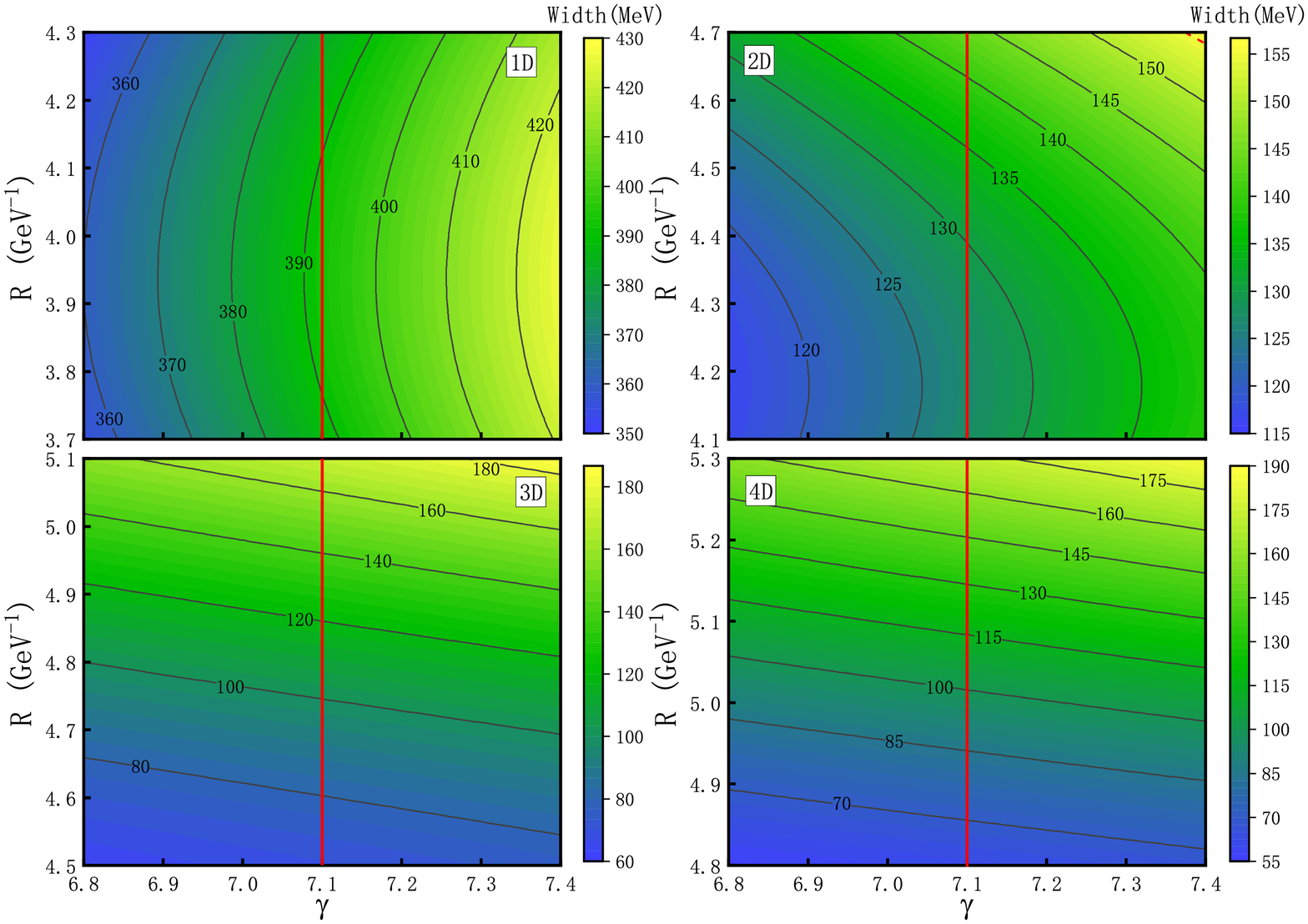}
  \caption{The total widths of $\rho_2$ family ($n=1$ to 4) with dependence of $R$ and $\gamma$. Different colors correspond to different width values and the contour curves are marked with the width values. Specially, the red longitudinal solid lines depicts the $R$ dependence when $\gamma=7.1$.}\label{rhocontour}
\end{figure*}

\begin{figure}[htbp]
  \centering
  \includegraphics[width=0.5\textwidth]{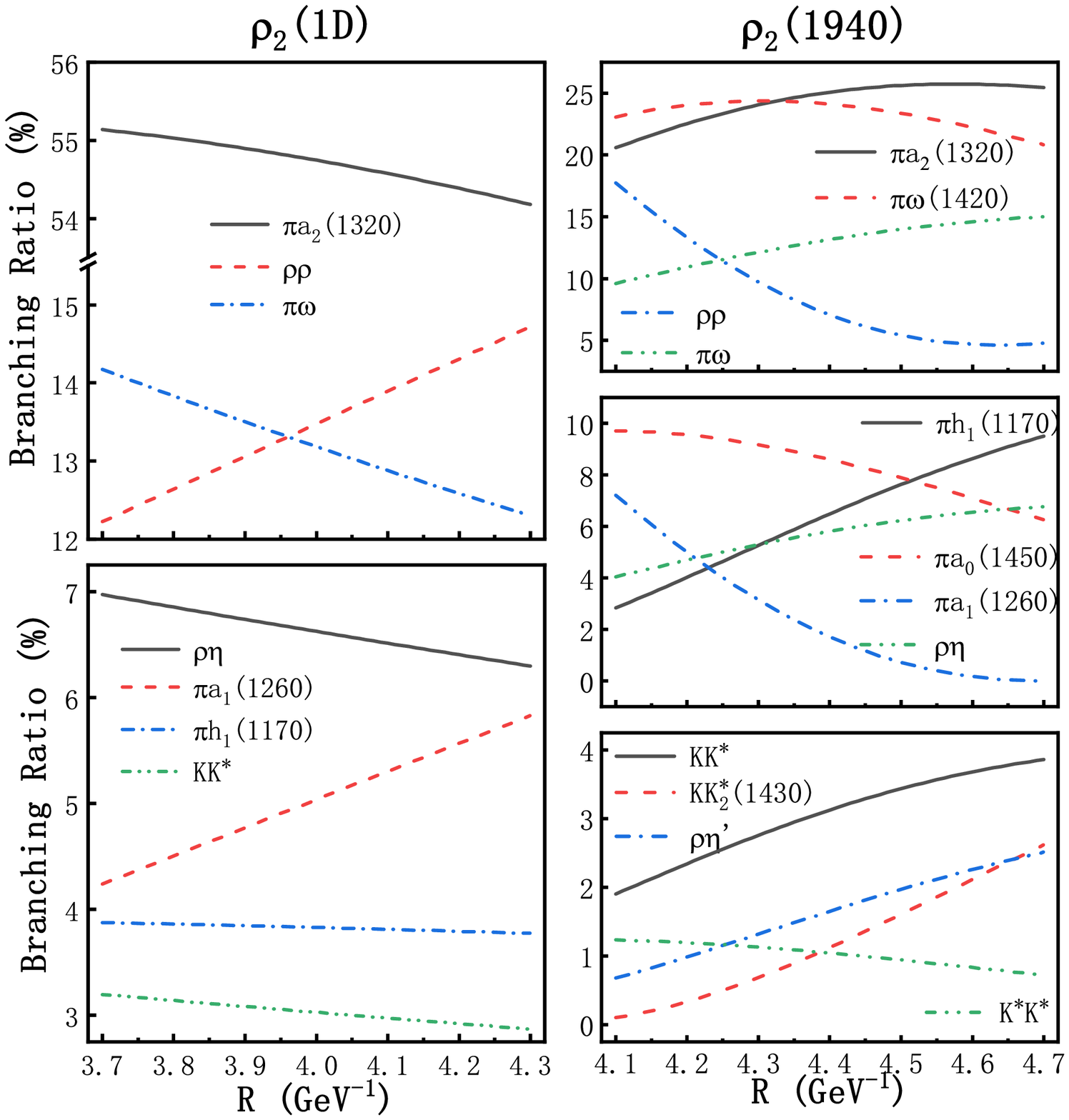}
  \caption{The branching ratios of $\rho_2(1D)$ and $\rho_2(1940)$ with $R$ dependence. Decay channels with branching ratios less than 1\% are neglected.}
  \label{rho12D}
\end{figure}

\begin{figure}[htbp]
  \centering
  \includegraphics[width=0.5\textwidth]{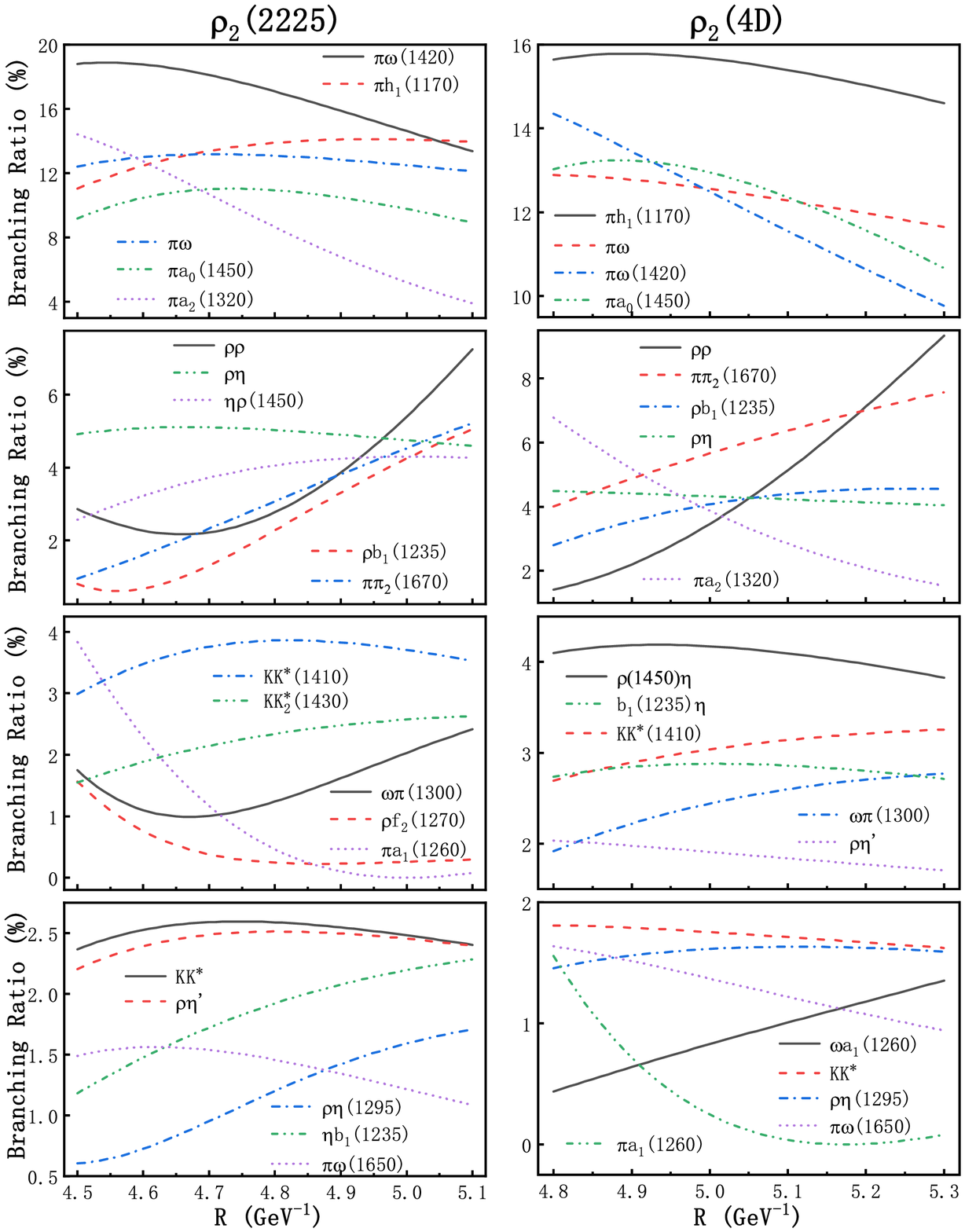}
  \caption{The branching ratios of $\rho_2(2225)$ and $\rho_2(4D)$ with $R$ dependence. Decay channels with branching ratios less than 1\% are neglected.}
  \label{rho34D}
\end{figure}

As the isovector counterpart of $\omega_2$ family, $\rho_2$ family takes the same $R$ ranges. The contour lines of total widths from $n$=1 to 4 are illustrated in Fig. \ref{rhocontour}.
Corresponding branching ratios are illustrated in Figs. \ref{rho12D} and \ref{rho34D}.
We can roughly conclude that the higher radial excitations rely more intensively on $R$ variation, and take larger $R$ values.

First, because no signal of $\rho_2(1D)$ is observed, a mass same with $\omega_2(1D)$ is obtained via GI model.
Shown as the top-left diagram of Fig. \ref{rhocontour}, the total width of $\rho_2(1D)$ is around 390 MeV when $\gamma$ = 7.1, consistent with Ref. \cite{Barnes:1996ff} but little less than the result of Ref. \cite{Godfrey:1998pd}.
From the left of Fig. \ref{rho12D},
the $\pi a_2(1320)$ mode of $\rho_2(1D)$ decay is predicted to obtain the largest branching ratio of $\sim$$55\%$. $a_2(1320)$ sequentially decays to $\rho\pi$, and $\rho_2(1D)$ eventually decays into $4\pi$ final state through this mode.
Furthermore, the secondary modes, $\rho\rho$, $\pi\omega$, $\pi a_1(1260)$, and $\pi h_1(1170)$ also contribute to $4\pi$ final state.
Therefore, why no signal of $\rho_2$ ground state is observed can be explained by that is difficult to find a broad width structure and complicated to reconstruct a $4\pi$ final state at experiment.
Though $\rho_2(1D)$ can be produced in principal, $\rho_2(1D)$ is hardly determined by Crystal Barrel detector because of quite low statistics.
Actually, the mass and width of $\rho_2(1940)$ already are poorly determined with low statistics for being close to the bottom of the available energy range.
However, $\rho_2(1D)$ also couples significantly to $\rho\eta$ and $KK^*$, so may be observable in both channels.
Besides, $\pi\omega$ mode should also be concerned since $\rho_2(1940)$ and $\rho_2(2225)$ are observed in this mode and the reconstruction efficiency is higher than $\omega\eta\pi$.
We also list some ratios weakly dependent on $R$ of partial decay width,  $\Gamma(\rho\rho)/\Gamma(\pi a_2(1320)) = 0.22-0.27$, $\Gamma(\pi\omega)/\Gamma(\pi a_2(1320)) = 0.23-0.26$, $\Gamma(\pi\omega)/\Gamma(\rho\rho) = 0.84-1.16$, $\Gamma(\rho\eta)/\Gamma(\pi\omega) = 0.49-0.51$, $\Gamma(\pi h_1(1170))/\Gamma(\rho\eta) = 0.56-0.60$, and $\Gamma(\rho\eta)/\Gamma(KK^*) = 2.18-2.19$.

According to mass spectrum analysis, the two resonances $\rho_2(1940)$ and $\rho_2(2225)$ observed in the processes $p\bar{p}\to\omega\pi^0$ and $p\bar{p}\to\omega\eta\pi^0$ \cite{Anisovich:2002su} are assigned as $2^3D_2$ and $3^3D_2$ states.
From the upper right of Fig. \ref{rhocontour}, $\rho_2(1940)$ has a total width range 127-144 MeV as $R$ varying from 4.1 to 4.7 $\mathrm{GeV}^{-1}$ when $\gamma$ = 7.1, and that well overlaps with experimental width $155\pm40$ MeV.
The strong two-body decay behavior of $\rho_2(1940)$ as $2^3D_2$ state is presented in the right of Fig.\ref{rho12D}. The main decay channels include $\pi a_2(1320)$, $\pi\omega(1420)$, and $\pi\omega$ with branching ratios 20.6-25.5\%, 20.8-23.1\%, and 9.6-15.0\%, respectively.
However, whether $\rho\rho$ is main channels depends on the $R$ value.
$\pi\omega(1420)$ also contributes to $\omega\eta\pi^0$ final state, and it makes sense $\rho_2(1940)$ discovered in $p\bar{p}\to\omega\eta\pi^0$ process.
Apart from channels contributing to $4\pi$ final state, $\rho_2(2D)$ obtains sizable branching ratios of kaons, $\rho\eta$ and $\rho\eta'$, and these modes can be carefully inspected at experiment.
Further information on typical ratios of partial decay widths are provided, $\Gamma(\pi a_2(1320))/\Gamma(\pi\omega(1420)) = 0.89-1.22$, $\Gamma(\pi a_2(1320))/\Gamma(\pi\omega) = 1.70-2.14$, $\Gamma(\rho\rho)/\Gamma(\pi\omega(1420)) = 0.23-0.76$, $\Gamma(\pi h_1(1170))/\Gamma(\pi a_2(1320)) = 0.14-0.37$, $\Gamma(\rho\eta)/\Gamma(\pi\omega) = 0.42-0.45$, $\Gamma(KK^*)/\Gamma(\rho\eta) = 0.47-0.57$, and $\Gamma(\rho\eta')/\Gamma(\rho\eta) = 0.17-0.37$.
These information are valuable for further experimental study and conformation of $\rho_2(1940)$.

Another $2^{--}$ resonance reported in Ref. \cite{Anisovich:2002su}, the $\rho_2(2225)$ as $3^3D_2$ state, is studied. Its partial decay width is presented in the left of Fig. \ref{rho34D}.
Since $\rho_2(2225)$ is assumed as the second excited state, its width is strongly dependent on $R$ value, and thus it possesses a broad width range with $R = 4.5-5.1 \mathrm{GeV^{-1}}$.
Unlike other three observed resonances, our calculated width is somewhat smaller than the experimental value $335^{+100}_{-50}$.
But notice that merely Crystal Barrel Collaboration announced this resonance and its experimental width error is evidently larger than other three resonances, so more experimental information is needed to testify our result.
The main decay channels of $\rho_2(2225)$ include $\pi\omega(1420)$, $\pi h_1(1170)$, $\pi\omega$, $\pi a_0(1450)$, and $\pi a_2(1320)$ with branching ratios equalling to 13.4-18.8\%, 11.0-14.1\%, 12.1-13.2\%, 8.9-11.0\%, and 3.9-14.4\%, respectively.
On the other hand, $\rho\rho$, $\rho\eta$, $\rho(1450)\eta$, $\rho b_1(1235)$, and $\pi\pi_2(1670)$ are subordinate decay channels.
Besides, sequential two-body decays of $\pi\omega(1420)$, $\omega a_2(1320)$, $\eta b_1(1235)$, and $\pi\omega(1650)$ channels give shares to $\omega\eta\pi^0$ final state, so it explains the experimental process $p\bar{p}\to\rho_2(2225)\to\omega\eta\pi^0$.
Here, detailed decay width ratios are also presented, $\Gamma(\pi\omega(1420))/\Gamma(\pi h_1(1170)) = 0.96-1.70$, $\Gamma(\pi\omega)/\Gamma(\pi\omega(1420)) = 1.10-1.51$, $\Gamma(\pi a_0(1450))/\Gamma(\pi\omega) = 0.74-0.84$, $\Gamma(\rho\eta)/\Gamma(\pi\omega) = 0.38-0.40$, $\Gamma(\rho(1450\eta))/\Gamma(\rho\eta) = 1.08-1.91$, $\Gamma(KK^*(1410))/\Gamma(KK^*_2(1430)) = 1.34-1.93$, $\Gamma(KK^*)/\Gamma(\rho\eta') = 1.00-1.07$, $\Gamma(\rho\eta')/\Gamma(\rho\eta) = 0.45-0.52$, and $\Gamma(\pi\omega(1650))/\Gamma(KK^*) = 0.45-0.63$.

\begin{figure*}[htbp]
  \centering
  \includegraphics[width=0.8\textwidth]{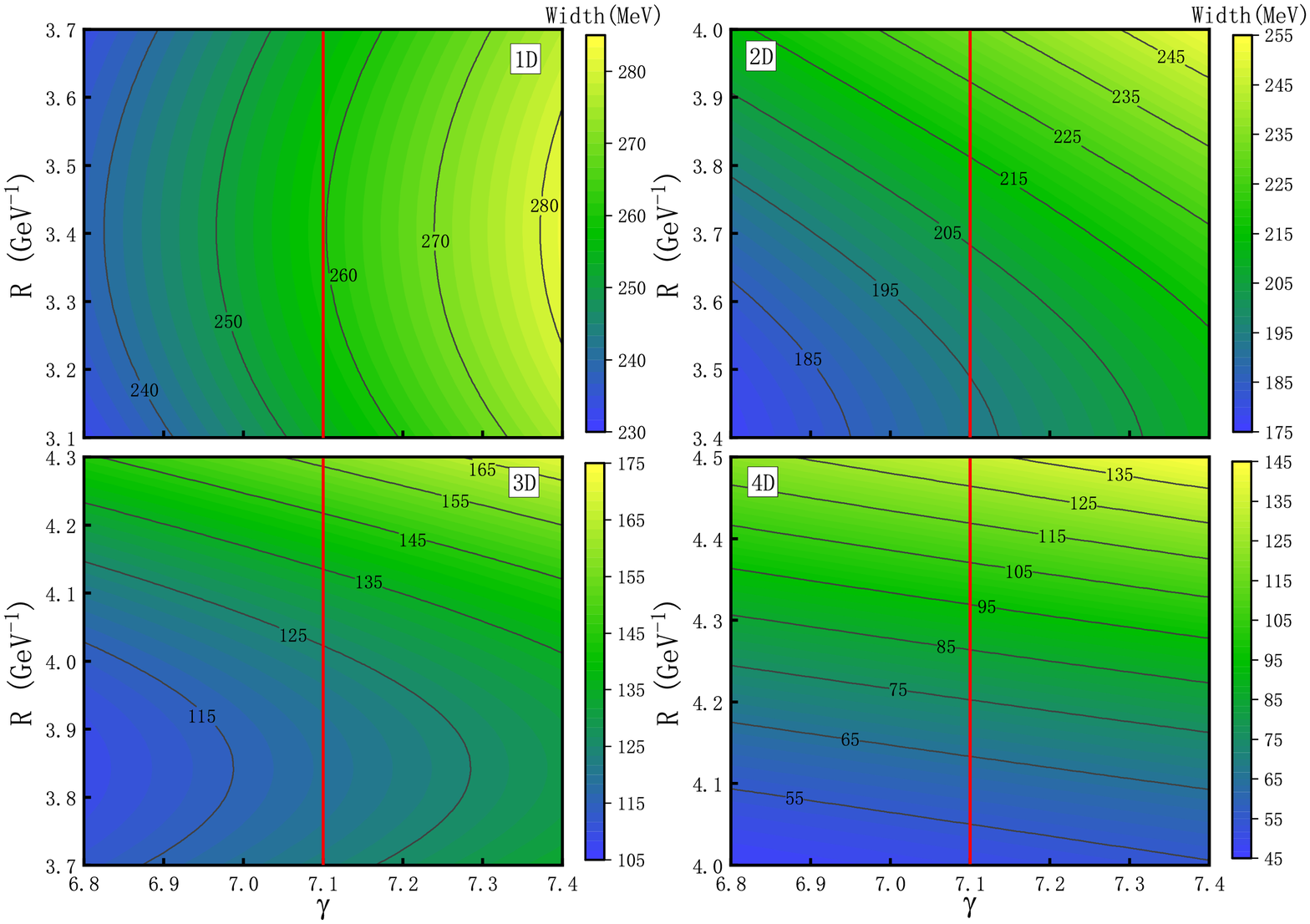}
  \caption{The total widths of $\phi_2$ family ($n=1$ to 4) with dependence of $R$ and $\gamma$. Different colors correspond to different width values and the contour curves are marked with the width values. Specially, the red longitudinal solid lines depicts the $R$ dependence when $\gamma=7.1$. }
  \label{phicontour}
\end{figure*}

The mass of $\rho_2(4D)$ absent at experiment is predicted as 2424 MeV, and its calculated total width is plotted in the lower right of Fig. \ref{rhocontour}, which presents a broad width range relying on $R$.
Our suggested total decaying width is 97 MeV when choosing suitable $R$ and $\gamma$ values.
The branching ratios of $\rho_2(4D)$ are illustrated in the right column of Fig. \ref{rho34D}.
In addition, because of higher excitation, both total width and the branching ratios are sharply dependent on $R$ value.
Main channels consist of $\pi h_1(1170)$, $\pi\omega$, $\pi\omega(1420)$ and $\pi a_0(1450)$.
Except $\pi h_1(1170)$ channel owning a stable ratio 15\%, other three main channels, $\pi\omega$, $\pi\omega(1420)$, and $\pi a_0(1450)$, show wider ratio ranges 11.6-12.9\%, 9.8-14.4\%, and 10.7-13.35\%, respectively.
Besides, we list several relative ratios weakly dependent on $R$, $\Gamma(\pi h_1(1170))\Gamma(\pi\omega) = 1.21-1.25$, $\Gamma(\pi\omega)/\Gamma(\pi\omega(1420)) = 0.90-1.19$, $\Gamma(\rho b_1(1235))/\Gamma(\rho\eta) = 0.62-1.13$, $\Gamma(\rho(1450)\eta)/\Gamma(\rho\eta) = 0.91-0.95$, $\Gamma(b_1(1235)\eta)/\Gamma(\rho(1450)\eta) = 0.67-0.71$, and $\Gamma(\rho\eta')/\Gamma(\rho\eta) = 0.42-0.45$,
These information is helpful for future experimental investigation.

%====================================================================================================================================
\subsection{$\phi_2$ Family}
%====================================================================================================================================

The information of $\phi_2$ family is still absent at experiment until now, so theoretical prediction of the absent states can assist experimentalist in searching for them and discovering more properties.
By combining the analysis of GI model and Regge trajectory, masses of $\phi_2(1D)-\phi_2(4D)$ are obtained in Table \ref{reggemass}, and their total widths are illustrated in Fig. \ref{phicontour}.
Considering our previous work of $K^*_2$ family \cite{Pang:2017dlw}, complete $J^{PC} = 2^{--}$ light meson nonets are established, and systematic analysis can be performed.
The analysis of $\phi_2$ states are proceeded in more detail as follows.

\begin{figure}[htbp]
  \centering
  \includegraphics[width=0.5\textwidth]{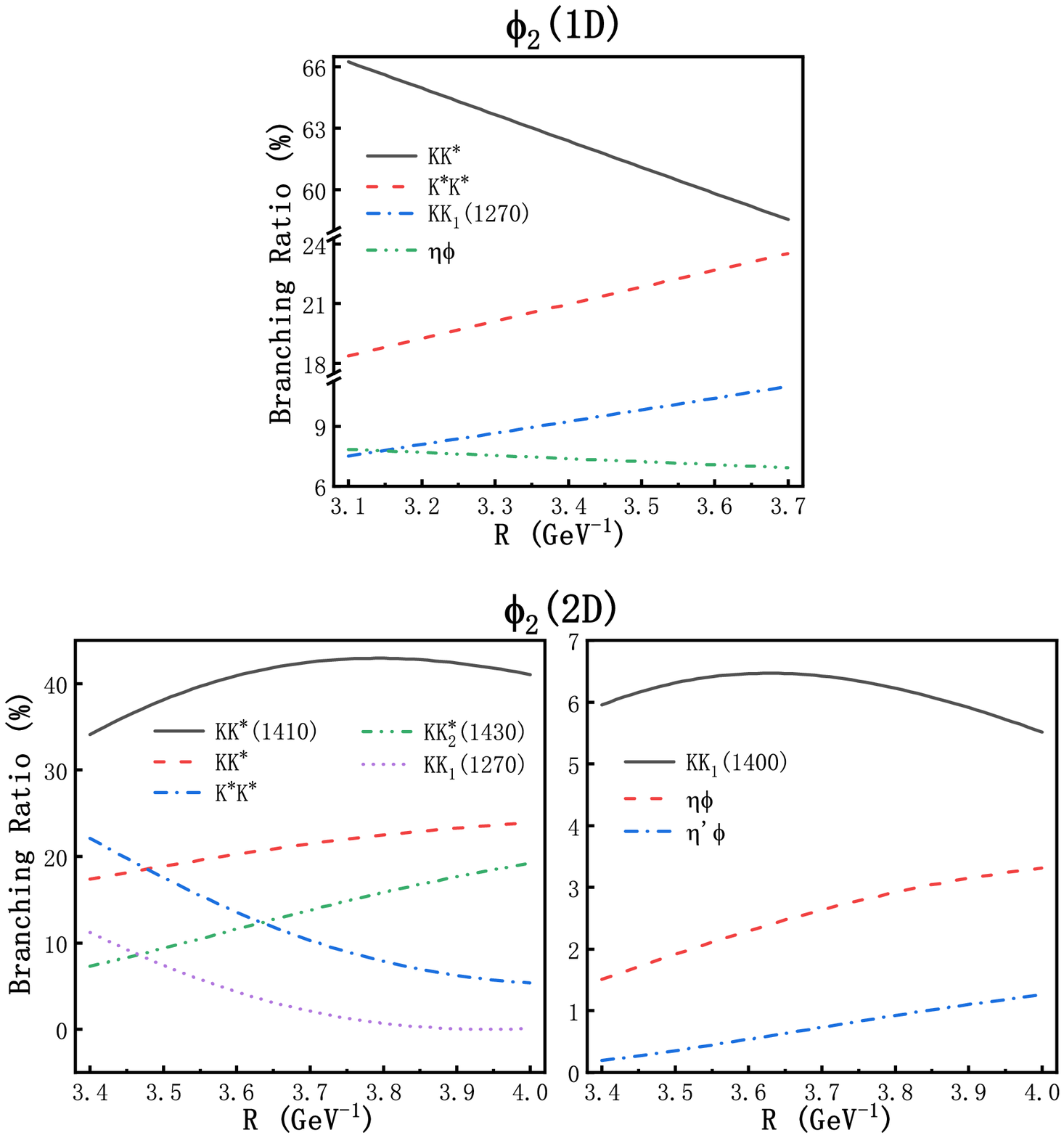}
  \caption{The branching ratios of $\phi_2(1D)$ and $\phi_2(2D)$ with $R$ dependence. Decay channels with branching ratios less than 1\% are neglected. }
  \label{phi12D}
\end{figure}

The mass of $\phi_2(1D)$ is suggested as 1904 MeV, near its partner $\phi_3(1850)$.
If taking a typical value of $\gamma = 7.1 $, its total width is predicted about 255 MeV with mild $R$ dependence shown in the top left of Fig. \ref{phicontour}, and it is comparable with the calculation in Ref. \cite{Godfrey:1998pd}.
Two-body decay properties of $\phi_2(1D)$ are presented at the top of Fig. \ref{phi12D}, where $KK^*$ decay mode is dominant with the fraction up to 58.5-66.2\%, and $K^*K^*$, $KK_1(1270)$, and $\eta\phi$ channels also hold considerable proportions.
Similar branching ratio results appear in Ref. \cite{Godfrey:1998pd}, except that of $\phi\eta$ is slightly larger than our prediction.
Besides, all channel ratios exhibit linearly dependence with changing $R$, which probably results from the smooth wavefunction of the ground state.
Thus, studying partial width ratios at experiment can appropriately test our prediction.
These are $\Gamma(K^*K^*)/\Gamma(KK^*) = 0.28-0.40$, $\Gamma(KK_1(1270))/\Gamma(K^*K^*) = 0.41-0.47$, and $\Gamma(KK_1(1270))/\Gamma(\eta\phi) = 0.96-1.59$.

Under the assignment of $2^3D_2$, the mass with 2151 MeV is employed, and corresponding total width and branching ratios are plotted in the top-right part of Fig. \ref{phicontour} and at the bottom of Fig. \ref{phi12D}, respectively.
The total width of $\phi_2(2D)$ is 206 MeV when taking the proposed $R$ value 3.7 $\mathrm{GeV^{-1}}$ and $\gamma = 7.1$.
The two-body strong decay properties, collected at the bottom of Fig. \ref{phi12D}, indicate that the decay channels devoting to main contributions include $KK^*(1410)$, $KK^*$, $K^*K^*$, $KK_2^*(1430)$, and $KK_1(1270)$, taking proportions 34.1-41.0\%, 17.4-23.0\%, 5.4-22.1\%, 7.3-19.2\%, and 0.1-11.2\%, respectively.
Since BESIII declared numerous state around 2.2 GeV, $\Gamma(\eta'\phi)/\Gamma(\eta\phi) = 0.19-0.38$ should be an ideal ratio to distinguish $\phi_2(2D)$.

\begin{figure}[htbp]
  \centering
  \includegraphics[width=0.5\textwidth]{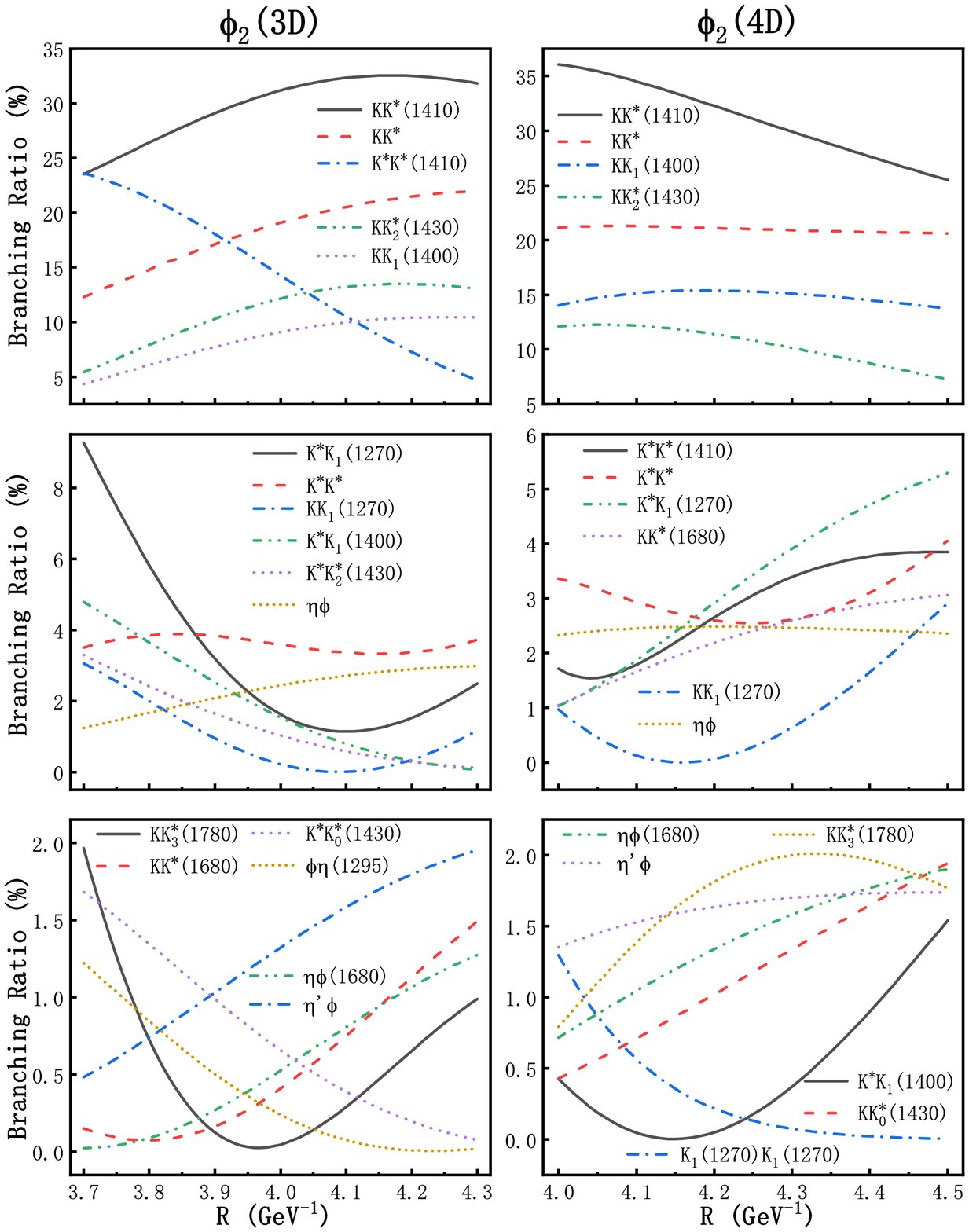}
  \caption{The branching ratios of $\phi_2(3D)$ and $\phi_2(4D)$ with $R$ dependence. Decay channels with branching ratios less than 1\% are neglected. }
  \label{phi34D}
\end{figure}

As for $\phi_2(3D)$, its mass is advised as 2372 MeV by the extrapolation from the Regge trajectory with the mass of the ground state calculated from the GI model.
Its recommended total width value is 124 MeV from the bottom left of Fig. \ref{phicontour}.
We list the branching ratios of $\phi_2(3D)$ in the left of Fig. \ref{phi34D}. $KK^*(1410)$, $KK^*$, $K^*K^*(1410)$, $KK_2^*(1430)$, $KK_1(1400)$, and $K^*K_1(1270)$ are main channels with branching ratios 23.5-31.8\%, 12.3-22.0\%, 23.6-4.7\%, 5.4-13.1\%, 4.3-10.4\%, and 1.1-9.3\%, respectively.
Especially, decay modes $K^*K^*(1410)$, $K^*K_1(1270)$, and $KK^*_3(1780)$ are significantly dependent on $R$ mainly because of the node effects.
By comparison, $KK^*(1410)$ is the dominant decay channel and more stable than other modes, and similar phenomenon appears for $\phi_2(2D)$ too.
Characteristic partial width ratio helpful for experimental exploration is $\Gamma(\eta'\phi)/\Gamma(\eta\phi) = 0.39-0.66$.

Finally, we analyze the decay behaviors of $\phi_2(4D)$ state with its mass proposed as 2574 MeV.
As the highest excitation considered in $\phi_2$ family, it provides complicated properties of strong decays due to the $R$ value sensitivity and node effects.
Its total width variation tendency is presented in the lower-right corner of Fig. \ref{phicontour} with advisable value 75 MeV when $\gamma$ = 7.1 and $R$ = 4.2 $\mathrm{GeV^{-1}}$.
Relevant branching ratios are shown in the right of Fig. \ref{phi34D}, where $KK^*(1410)$ is the dominant mode holding 25.5-36.0\% fraction, and $KK^*$, $KK_1(1400)$, and $KK^*_2(1430)$ are also main channels.
Some typical relative branching fractions are presented here, $\Gamma(KK_1(1400))/\Gamma(KK^*) = 0.66-0.73$, $\Gamma(\eta\phi)/\Gamma(K^*K^*) = 0.58-0.97$, and $\Gamma(\eta'\phi)/\Gamma(\eta\phi) = 0.58-0.74$.
These predicted information of the missing $\phi_2$ family may inspire more experimental efforts for hunting them.

%====================================================================================================================================
\section{summary and discussion}\label{sec4}
%====================================================================================================================================

Since the quark model proposed, it dose achieve great success for explaining and predicting numerous hadrons.
However, there also remain some unexplained exotic states beyond the quark model, such as glueballs, hybrids.
To distinguish between conventional and exotic mesons, it is crucial that we comprehend conventional-meson spectroscopy very well.
Currently, the experimental information of $J^{PC}=2^{--}$ unflavored light mesons is still scarce except the results of Crystal Barrel detector in 2002, and thus they are not well established.
Therefore, a systematical and complete study of $2^{--}$ unflavored light mesons is necessary to stimulate mounting experimental efforts to this sector.

In this work, an investigation of mass spectrum is firstly performed by combining the analysis of Regge trajectory and GI model in Sec. \ref{sec2}, which indicates that the four experimental resonances are assigned as the first and second excitation of $\omega_2$ and $\rho_2$ families, respectively.
In addition, this categorization is also supported by the corresponding features of two-body decays allowed by OZI rule.
More importantly, with predicted masses $\sim$1.7 GeV for $\omega_2$ and $\rho_2$ ground states, total decay widths are suggested as 220 MeV and 390 MeV.
Furthermore, since $\omega_2(1D)$ is masked by $\pi_2(1670)$ and $\rho_2(1D)$ is difficult to reconstruct with broad width and $4\pi$ final state, we can explain why they are blind at experiment.
Detailed decay properties of the missing ground resonances are given by QPC model in Sec. \ref{sec3}.
In addition, for the absolutely missing $\phi_2$ family, their partial decay behaviors are also discussed by adopting advisable masses from $n$ = 1 to 4.
The branching ratios of main and subordinate channels are illustrated in Sec. \ref{sec3}, respectively.
Certain partial decay width is obtained by multiplying the corresponding branching ratio with total width.
Definitely, the main decay widths of ground states are listed in Table \ref{groundwidth}.

\begin{table}[htbp]
  \centering
  \caption{The partial decay width of ground states in units of MeV. Corresponding $R$ value ranges are 3.7-4.3 for $\omega_2$ and $\rho_2$ ground states, and 3.1-3.7 for $\phi_2$ ground state.}
  \label{groundwidth}
  \begin{tabular}{lc|lc|lc}
     \hline\hline
     Channels        &$\omega_2(1D)$&Channels       &$\rho_2(1D)$& Channels   &$\phi_2(1D)$ \\
     \hline
     $\pi\rho$       &142-165       &$\pi a_2(1320)$&206-213     &$KK^*$      &149-167      \\
     $\pi b_1(1235)$ &25-26         &$\rho\rho$     &47-56       &$K^*K^*$    &46-59        \\
     $\omega\eta$    &24-27         &$\pi\omega$    &47-55       &$KK_1(1270)$&19-28      \\
     $KK^*$          &11-12         &$\rho\eta$     &24-27       &$\eta\phi$  &18-20   \\
                     &              &$\pi a_1(1260)$&16-22       &            &  \\
     \hline
     \hline
   \end{tabular}

\end{table}

Owning to merely theoretical inputs of the masses for most states, it is necessary to discuss the mass dependence of decay widths.
For example, when adding 50 MeV to the masses of $\omega_2$ family, main decay widths of $\omega_2(4D)$ increase by 24\% maximumly.
As to lower state, their main decay widths are relatively less sensitive to mass variation.
Similarly, mass dependence of $\rho_2$ and $\phi_2$ families can be estimated.

The strong decay information is vital to search for missing ground states and higher excitations.
We hope our study can inspire more experimental interests to find them and establish $2^{--}$ light meson family at experiment eventually.
BESIII, CMD-3, SND, KEDR, COMPASS, and E852 are suitable platforms to explore these issues.
Especially, BESIII is the promising facility with the highest luminosity running in 2.0-4.6 GeV, and its main focus still is exploring light hadrons in the following years.
Besides, it is hopeful that many states can be experimentally better understood in relatively near future at the Jefferson Lab.
Potential channels, $\omega\eta$ for $\omega_2$, $\rho\eta$ for $\rho_2$, $\eta\phi$ for $\phi_2$, are suggested to explore these ground states.
$KK^*$ is an ideal channel which contributes to all $2^{--}$ light unflavored states, and contributes to sizable branching ratios.

%====================================================================================================================================
\section{acknowledgement}\label{sec4}
%====================================================================================================================================
This project is supported by the China National Funds for Distinguished Young Scientists under Grant No. 11825503. This project is also supported by the National Natural Science Foundation of China under Grant No. 11705072 and Natural Science Foundation Project of Qinghai Office of Science and Technology (No. 2017-ZJ-748).

%====================================================================================================================================

\end{document}